\documentclass[pre,twocolumn,showpacs]{revtex4}

\usepackage{amssymb,amsmath}

\usepackage{graphicx}

\begin{document}
\title{Transport coefficients of driven granular fluids at moderate volume fractions}
\author{Vicente Garz\'o}
\email{vicenteg@unex.es} \homepage{http://www1.unex.es/eweb/fisteor/vicente/} \affiliation{Departamento de
F\'{\i}sica, Universidad de Extremadura, E-06071 Badajoz, Spain}

\begin{abstract}

In a recent publication [Phys. Rev. E \textbf{83}, 011301 (2011)], Vollmayr--Lee \emph{et al.}  have determined by computer simulations the thermal diffusivity and the longitudinal viscosity coefficients of a driven granular fluid of hard spheres at intermediate volume fractions. Although they compare their simulation results with the predictions of kinetic theory, they use the dilute expressions for the driven system and the modified Sonine approximations for the undriven system. The goal here is to carry out this comparison by proposing a modified Sonine approximation to the Enskog equation for driven granular fluids that leads to a better quantitative agreement.

\end{abstract}

\draft \pacs{05.20.Dd, 45.70.Mg, 51.10.+y, 05.60.-k}

\date{\today}
\maketitle

In a recent paper, Vollmayr-Lee, Aspelmeier, and Zippelius \cite{VAZ11} have determined the time-delayed correlation functions of a homogeneous granular fluid of hard spheres at intermediate volume fractions. As usually done in computer simulations, the fluid is driven by means of a stochastic external force such that the particles are randomly kicked with a given frequency \cite{WM96}. In the steady state, where the external energy input due to driving compensates for collisional cooling, they measured the dynamic structure factor $S(q,\omega)$ ($q$ is the wave number and $\omega$ is the angular frequency) for several values of the coefficient of restitution $\epsilon$ and volume fractions $\eta$. The corresponding best fit of the simulation results of $S(q,\omega)$ allowed them to identify (for the smallest values of $q$) the thermal diffusivity $D_T$ and the longitudinal viscosity $\nu_1$ coefficients. These transport coefficients are defined as
\begin{equation}
\label{1}
D_T=\frac{2}{dn}\kappa,
\end{equation}
\begin{equation}
\label{2}
\nu_1=\frac{1}{mn}\left(2\frac{d-1}{d}\eta_{\text{shear}}+\zeta\right),
\end{equation}
where $n$ is the number density, $m$ is the mass of a particle, $d$ is the dimensionality of the system, $\kappa$ is the thermal conductivity, $\eta_{\text{shear}}$ is the shear viscosity and $\zeta$ is the bulk viscosity. The authors compared their simulation results for $D_T$ and $\nu_1$ with theoretical predictions of kinetic theory (see Table II of Ref.\ \cite{VAZ11}) and concluded that the latter agree well with computer simulations. In this paper, Vollmayr--Lee \emph{et al.} \cite{VAZ11} used three different analytical results, two of them \cite{GM02,GSM07} obtained from the revised Enskog theory (RET) and one of them obtained from a (simple) kinetic model of the RET \cite{DBS97}. However, the kinetic theory expressions for the transport coefficients $\kappa(\eta)$ and $\eta_{\text{shear}}(\eta)$ employed for making the comparison (namely, those derived from the true RET\cite{GM02,GSM07} and not from a simple model of the latter equation \cite{DBS97}) correspond actually to their low-density forms and so they do not incorporate the density corrections accounted for by the Enskog kinetic theory \cite{note}. The question arises then as to whether, and if so to what extent, the conclusions drawn from Ref.\ \cite{VAZ11} may be altered when the Enskog expressions for $D_T$ and $\nu_1$ are considered. This question is addressed here and additionally, a theory based on the Enskog equation is proposed for driven systems that leads to a better qualitative agreement with simulation data.

We consider a granular fluid composed of smooth inelastic disks ($d=2$) or spheres ($d=3$) of mass $m$ and diameter $a$. Collisions are characterized by a (constant) coefficient of normal restitution $0<\epsilon \leq 1$, with $\epsilon=1$ in the elastic limit. In the hydrodynamic regime, the Navier-Stokes constitutive equations of the pressure tensor $P_{ij}$ and the heat flux ${\bf q}$ are given by
\begin{equation}
\label{3}
P_{ij}=p \delta_{ij}-\eta_{\text{shear}}\left(\nabla_j u_i+\nabla_i u_j-\frac{2}{d}\delta_{ij}\nabla \cdot {\bf u}\right)-\zeta \delta_{ij}\nabla \cdot {\bf u},
\end{equation}
\begin{equation}
\label{4}
{\bf q}=-\kappa \nabla T-\mu \nabla n,
\end{equation}
where $p$ is the hydrostatic pressure, ${\bf u}$ is the flow velocity, and $T$ is the granular temperature. Moreover, apart from the shear and bulk viscosities and the thermal conductivity coefficient, the coefficient $\mu$ is a new transport coefficient not present in the elastic case. Explicit expressions for the above transport coefficients can be obtained from kinetic theory. In particular, the RET provides a reliable description of transport properties of a granular fluid at moderate densities and across a wide range of degrees of dissipation. The RET has been solved by applying the well-known Chapman-Enskog method \cite{CC70} conveniently adapted to dissipative dynamics. First attempts \cite{LSJC84,JR85,SG98} to determine the Enskog transport coefficients were carried out for nearly elastic particles so that the results only hold in principle in the quasielastic limit ($\epsilon \lesssim 1$). Subsequent works \cite{GD99,L05} based on the Chapman-Enskog method do not impose any constraints on the level of dissipation and take into account the (complete) nonlinear dependence of the transport coefficients on the coefficient of restitution $\epsilon$. On the other hand, as for elastic collisions \cite{CC70}, the transport coefficients are given in terms of the solutions of a set of coupled linear integral equations that are solved by means of a polynomial Sonine expansion. For simplicity, usually only the lowest Sonine polynomial (first Sonine approximation) is retained \cite{GD99,L05} and the results obtained from this simple approach agrees well with Monte Carlo simulations, except at high dissipation for the heat flux transport coefficients \cite{BM04}. Motivated by this disagreement, a modified version of the first Sonine approximation was proposed by Garz\'o \emph{et al.} \cite{GSM07}. This modified Sonine approach significantly improves the $\epsilon$-dependence of $\kappa$ and $\mu$ and corrects the discrepancies between simulation and theory for a dilute gas. The theory proposed in Ref.\ \cite{GSM07} (in its dilute version) was employed in the comparison carried out  by Vollmayr-Lee \emph{et al.} \cite{VAZ11} in Table II of their paper.

An important point in all the above theories \cite{GD99,L05,GSM07} is that they have been derived from an expansion around the free cooling state, namely, a reference state that depends on time through its dependence on granular temperature. On the other hand, as said before, many of the computer simulations (as the one carried out in Ref.\ \cite{VAZ11}) are performed by using stochastic driving mechanisms to maintain the system in a stationary state. Given that this external energy input does not play a neutral role on the dynamics properties of the fluid \cite{GS03}, the transport coefficients in the undriven (unforced) case \cite{GD99,L05,GSM07} differ from those obtained in the presence of an external thermostat. In particular, the coefficients $\eta_{\text{shear}}$, $\zeta$, $\kappa$, and $\mu$ were determined in the Appendix B of Ref.\ \cite{GM02} for a \emph{dense} fluid of hard spheres ($d=3$) heated by a stochastic thermostat. However, as in Refs.\ \cite{GD99,L05}, the results were obtained by considering the (standard) first Sonine approximation. To the best of my knowledge, the corresponding modified version to the driven case has not been derived so far.
One of the goals of the present paper is to implement this modified version of the first Sonine approximation when the fluid is driven by means of the stochastic thermostat. Given that the derivation of this theory follows similar mathematical steps as those made in the undriven case \cite{GSM07}, here only the final expressions for the transport coefficients $\eta_{\text{shear}}$, $\zeta$, and $\kappa$ will be displayed.
\begin{table}[tbp]
\caption{Comparison of theoretical predictions and computer simulations}
\label{table1}
\begin{ruledtabular}
\begin{tabular}{ccccc}
& \multicolumn{2}{c}{$\epsilon=0.8$} & \multicolumn{2}{c}{$\epsilon=0.9$}
\\
\hline
$\eta=0.05$& $D_T$ & $\nu_1$ & $D_T$ & $\nu_1$\\
\hline
MD results& 4.72 & 2.55 & 4.63 & 3.23 \\
Undriven RET&5.99&2.86&5.60&2.80\\
Driven RET&3.88&2.65&4.44&2.69\\
\hline
$\eta=0.1$& $D_T$ & $\nu_1$ & $D_T$ & $\nu_1$\\
\hline
MD results& 2.23 & 1.20 & 2.67 & 1.70 \\
Undriven RET&3.35&1.71&3.27&1.73\\
Driven RET&2.22&1.60&2.62&1.67\\
\hline
$\eta=0.2$& $D_T$ & $\nu_1$ & $D_T$ & $\nu_1$\\
\hline
MD results& 1.95 & 1.02 & 2.22 & 1.10 \\
Undriven RET&2.45&1.64&2.58&1.73\\
Driven RET&1.74&1.58&2.15&1.69
\end{tabular}
\end{ruledtabular}
\end{table}

Since the modified Sonine approximations applied to the RET for the undriven \cite{GSM07} and driven cases are perhaps the two most accurate approaches for the Enskog transport coefficients, only both theories are used in this paper. Moreover, for the sake of comparison, the same units as those taken in Ref.\ \cite{VAZ11} ($m=a=T=1$) are chosen. In a compact form, the coefficients $\eta_{\text{shear}}$, $\zeta$, and $\kappa$ can be written  as
\begin{equation}
\label{5}
\zeta(\epsilon,\eta)=\frac{\Gamma \left(\frac{d}{2}\right)}{\pi^{(d+1)/2}}
2^{2(d-1)}\eta^2 \chi
(1+\epsilon)\left(1-\frac{c}{32}\right),
\end{equation}
\begin{eqnarray}
\label{6}
\eta_{\text{shear}}(\epsilon,\eta)&=& \frac{\Gamma \left(\frac{d}{2}\right)}{8\pi^{(d-1)/2}}
\frac{d+2}{\nu_{\eta}-b_\eta\xi}\bigg[1-\frac{2^{d-2}}{d+2}
(1+\epsilon)\nonumber\\
& &
\times(1-3 \epsilon)\eta \chi\bigg]\bigg[1+\frac{2^{d-1}}{d+2}\eta \chi(1+\epsilon)\bigg]\nonumber\\
& & +\frac{d}{d+2}\zeta,
\end{eqnarray}
\vspace{0.2mm}
\begin{eqnarray}
\label{7}
\kappa(\epsilon,\eta)&=&\frac{\Gamma \left(\frac{d}{2}\right)}{16\pi^{(d-1)/2}}
\frac{(d+2)^2}{\nu_{\kappa}-b_\kappa\xi}
\left\{1+c+3\frac{
2^{d-3}}{d+2}\eta\chi(1+\epsilon)^2\right.\nonumber\\
&\times & \left.\left[\frac{1+\epsilon}{2}c-1+2\epsilon\right]\right\}\left[1+
3\frac{2^{d-2}}{d+2}\eta \chi
(1+\epsilon)\right]\nonumber\\
&+&\frac{d\Gamma \left(\frac{d}{2}\right)}{\pi^{(d+1)/2}}
2^{2d-3}  \eta^2 \chi
(1+\epsilon)\left(1+\frac{7}{32} c \right),
\end{eqnarray}
where
\begin{equation}
\label{8}
\nu_\eta=\frac{3}{4d}\chi \left(1-\epsilon+\frac{2}{3}d\right)(1+\epsilon)
\left(1+\frac{7}{32}c\right),
\end{equation}
\begin{eqnarray}
\label{9}
\nu_\kappa&=&
\frac{1+\epsilon}{d}\chi\left[\frac{d-1}{2}+\frac{3}{16}(d+8)
(1-\epsilon)\right.\nonumber\\
&&\left.+\frac{296+217d-3(160+11d)\epsilon}{512}c\right].
\end{eqnarray}
In the above equations, $\eta=[\pi^{d/2}/2^{d-1}d\Gamma(d/2)]n$ is the volume fraction and $\chi(\eta)$ is the pair correlation function at contact. In addition, $b_\eta=b_\kappa=0$ and
\begin{equation}
\label{10}
c(\epsilon)=\frac{32(1-\epsilon)(1-2\epsilon^2)}{73+56d-3\epsilon(35+8d)+30(1-\epsilon)\epsilon^2},
\end{equation}
in the case of the \emph{driven} (stochastic thermostat) RET.  In the case of the \emph{undriven} RET \cite{GSM07}, one has $b_\eta=\frac{1}{2}$, $b_\kappa=2$,
\begin{equation}
\label{11}
c(\epsilon)=\frac{32(1-\epsilon)(1-2\epsilon^2)}{25+24d-\epsilon(57-8d)-2(1-\epsilon)\epsilon^2},
\end{equation}
and
\begin{equation}
\label{12}
\xi=\frac{d+2}{4d}(1-\epsilon^2)\chi\left(1+\frac{3}{32}c\right).
\end{equation}
\begin{figure}
\includegraphics[width=0.8 \columnwidth,angle=0]{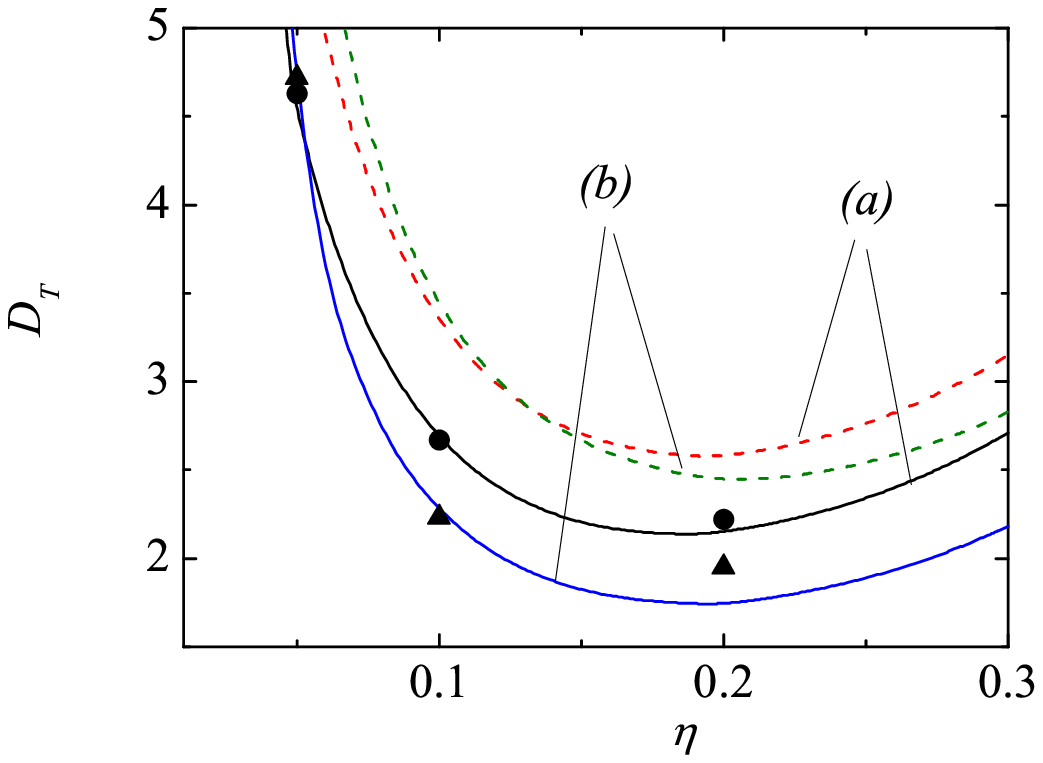}
\caption{(color online) Plot of the thermal diffusivity $D_T$ as a function of the volume fraction $\eta$ for two different values of the coefficient of restitution $\epsilon$: $\epsilon =0.9$ (a) and $\epsilon =0.8$ (b). The solid and dashed lines are the theoretical predictions derived for the driven and undriven gas, respectively. Symbols are simulation results obtained by Vollmayr--Lee {\em et al.} \cite{VAZ11} at $\epsilon=0.9$ (circles) and at $\epsilon=0.8$ (triangles).
\label{fig2}}
\end{figure}

\begin{figure}
\includegraphics[width=0.8 \columnwidth,angle=0]{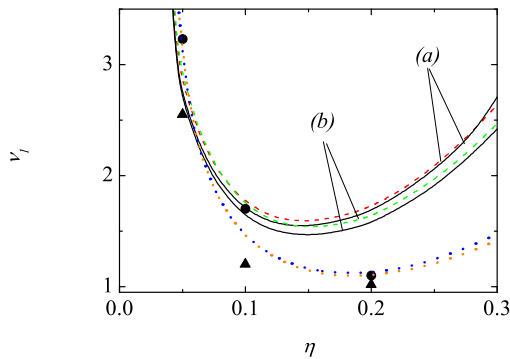}
\caption{(color online) Plot of the longitudinal viscosity $\nu_1$ as a function of the volume fraction $\eta$ for two different values of the coefficient of restitution $\epsilon$: $\epsilon =0.9$ (a) and $\epsilon =0.8$ (b). The solid and dashed lines are the theoretical predictions derived for the driven and undriven gas, respectively. Symbols are simulation results obtained by Vollmayr--Lee {\em et al.} \cite{VAZ11} at $\epsilon=0.9$ (circles) and at $\epsilon=0.8$ (triangles). The dotted lines are the theoretical results (blue for $\alpha=0.9$ and orange for $\alpha=0.8$) obtained from the driven theory but when one considers the low density value of the shear viscosity coefficient.
\label{fig4}}
\end{figure}

For hard spheres ($d=3$), as in Ref.\ \cite{VAZ11}, the Carnahan-Starling approximation \cite{CS69} for $\chi$ is considered:
\begin{equation}
\label{13}
\chi=\frac{1-\frac{\eta}{2}}{(1-\eta)^3}.
\end{equation}
In the three-dimensional case, according to Eqs.\ \eqref{1} and \eqref{2}, the thermal diffusivity $D_T$ and the longitudinal viscosity $\nu_1$ are defined as $D_T=(\pi/9\eta)\kappa$ and $\nu_1=(\pi/6\eta)[(4/3)\eta_{\text{shear}}+\zeta]$.

Table \ref{table1} reports the comparison between molecular dynamics (MD) results and theoretical predictions given by the RET in the undriven \cite{GSM07} and driven cases for hard spheres. Given that the fits of the simulation results require $q$-dependent transport coefficients, here only the fit result for the smallest value of the wave number $q$ is considered. An inspection of the results displayed in the table shows that both kinetic theory results (undriven and driven RET) reproduce the trends of simulation data, except perhaps in the case of the longitudinal viscosity $\nu_1$ at the largest volume fraction considered ($\eta=0.2$). At a more quantitative level, it is apparent that the analytical results derived in the presence of the stochastic thermostat agree better with MD than those obtained in the unforced case. This is the expected result since the simulations were carried out under the action of this driving mechanism. It is especially significant the good agreement found for the thermal diffusivity coefficient $D_T$, where the discrepancies between theory and simulation are less than 10\%, even for strong dissipation ($\alpha=0.8$). It must be noted that the results derived for $D_T$ in the undriven case from the kinetic model of the RET \cite{DBS97} agree better with simulation data than those obtained from the true Enskog equation for an undriven granular gas\cite{GSM07}. This is essentially because the impact of dissipation on $D_T$ predicted by the kinetic model is less significant than the one obtained from the RET for undriven systems, which agrees qualitatively with MD simulations of driven fluids \cite{VAZ11}.   

In order to illustrate the complete $\eta$-dependence of the above transport coefficients, Figs.\ \ref{fig2} and \ref{fig4} show $D_T$ and $\nu_1$ versus $\eta$, respectively, for $\alpha=0.9$ and $0.8$.
As said before, we observe that in general the influence of dissipation on the thermal diffusivity $D_T$ in the driven case is more important than in the undriven case, which is consistent with MD data. The influence of the thermostat on the longitudinal viscosity $\nu_1$ is less significant than for $D_T$ since both theories agree very well. On the other hand, the driven RET is still superior to the undriven RET, although the former overestimates the simulation data for moderate densities. Surprisingly, the comparison agrees better when one uses the \emph{dilute} shear viscosity form $\eta_{\text{shear}}(0)$ instead of the corresponding Enskog coefficient $\eta_{\text{shear}}(\eta)$ for the coefficient $\nu_1(\eta)$. This is clearly shown in Fig.\ \ref{fig4} where the combination $(\pi/6\eta)[(4/3)\eta_{\text{shear}}(0)+\zeta(\eta)]$ is also plotted for comparison. The disagreement between the driven RET and MD at moderate densities could be due to the fact that the density dependence of the shear viscosity is not well captured by the \emph{modified} Sonine solution to the Enskog equation or this can also reflect the limitations of the RET as the granular fluid becomes denser. In this latter case (strong dissipation and moderate densities), velocity correlations among the particles which are about to collide (which are absent in the Enskog description) could play a significant role in the dynamics of the system \cite{ML98}. On the other hand, more comparisons between the results derived from the Enskog equation and computer simulations are needed before qualitative conclusions can be drawn.

To summarize, I have revisited the comparison for the thermal diffusivity $D_T$ and the longitudinal viscosity $\nu_1$ carried out in Ref.\ \cite{VAZ11} between kinetic theory and MD simulations for driven granular fluids. As Eqs.\ \eqref{1} and \eqref{2} show, $D_T$ and $\nu_1$ are defined in terms of
the shear and bulk viscosities and the thermal conductivity coefficient. Although the simulations
of Ref.\ \cite{VAZ11} were performed at \emph{moderate} densities, in order to compute $D_T$ and $\nu_1$ Vollmayr-Lee  \emph{et al.}\cite{VAZ11} used the \emph{dilute} expressions of the shear viscosity and the thermal conductivity coefficients instead of their corresponding Enskog forms. The present comparison (displayed in Table \ref{table1} and in Figs.\ \ref{fig2} and \ref{fig4}) shows that in general the resulting transport coefficients obtained from the RET in the driven case (by using a modified Sonine approximation) agree better with computer simulations (which were carried out by introducing a stochastic driving mechanism) than those derived in the undriven case \cite{GSM07}. This contrasts with the comparison made in Ref.\ \cite{VAZ11} where the results obtained for an unforced gas were shown to be superior to the other ones.
On the other hand, at a more quantitative level, while the driven theory compares very well with simulation data for $D_T$ in the wide range of densities, some discrepancies (see Fig.\ \ref{fig4}) appear in the case of $\nu_1$ as the gas becomes denser. These discrepancies can be mitigated if one considers the dilute form ($\eta=0$) for the shear viscosity $\eta_{\text{shear}}$ in the definition \eqref{2} of $\nu_1$. This suggests that perhaps the dependence of $\eta_{\text{shear}}$ on the density $\eta$ is less important than the one predicted by the Enskog equation and consequently, the $\eta$-dependence of $\nu_1$ is essentially given by the bulk viscosity $\zeta$ [which is well captured by its Enskog form \eqref{5}].

Before closing this paper, let me offer some remarks on the comparison made here between kinetic theory and MD simulations. As in the case of ordinary (elastic) fluids, the Enskog kinetic equation takes into account spatial correlations through the pair correlation function but neglects velocity correlations (molecular chaos hypothesis). In this context, the Enskog equation can be considered as an accurate and practical generalization of the Boltzmann equation to finite densities. On the other hand, although the molecular chaos assumption can be questionable as the density of granular fluid increases \cite{ML98}, there is still substantial evidence in the literature on the validity of the Enskog theory for densities outside the Boltzmann limit and values of dissipation beyond the quasielastic limit \cite{Enskog}. The results presented here give support again to the Enskog theory as a reliable basis for the description of dynamics across a wide range of densities, length scales, and degrees of dissipation.

\acknowledgments
The present work has been supported by the Ministerio de
Educaci\'on y Ciencia (Spain) through grant No. FIS2010-16587, partially financed by
FEDER funds and by the Junta de Extremadura (Spain) through Grant No. GRU10158.


\begin{thebibliography} {99}
\bibitem{note}There is a misprint in the first line of Eq. (B8) of Ref.\ \cite{GM02} since the combination $1+[1+c(\cdots)]$ should be $1+c(\cdots)$. With this change, Eq. (B8) is consistent with Eq. (47) of Ref.\ \cite{GM02} in the low density limit ($\eta\to 0$).
\end{thebibliography}
\end{document}